\documentclass[12pt, a4paper, oneside]{article}
\usepackage[a4paper,left=2.5cm,right=2.5cm,top=2.5cm,bottom=2.5cm]{geometry}
\usepackage{amssymb}
\usepackage{amsmath}
\usepackage{graphicx}
\usepackage{authblk}
\usepackage{subcaption}
\begin{document}
\title{Numerical Calculation of Hubble Hierarchy Parameters and Observational Parameters of Inflation}

\author[1]{Milan Milo\v sevi\'c\thanks{mmilan@seenet-mtp.info}}
\author[2]{Neven Bili\'c}
\author[1]{Dragoljub D. Dimitrijevi\' c}
\author[1]{Goran S. Djordjevi\' c}
\author[3]{Marko Stojanovi\' c}
\affil[1]{Department of Physics, Faculty of Sciences and Mathematics, University of Ni\v s, Serbia}
\affil[2]{Division of Theoretical Physics, Rudjer Bo\v{s}kovi\'{c} Institute, Zagreb, Croatia}
\affil[2]{Faculty of Medicine, University of Ni\v s,  Serbia}

\maketitle

\begin{abstract}
We present results obtained by a software we developed for computing observational cosmological inflation parameters: the scalar spectral index ($n_s$) and the tensor-to-scalar ratio ($r$) for a standard single field and tachyon inflation, as well as for a tachyon inflation in the second Randall-Sundrum model with an additional radion field. The calculated numerical values of observational parameters are compared with the latest results of observations obtained by the Planck Collaboration. The program is written in C/C++. The \textit{GNU Scientific Library} is used for some of the numerical computations and R language is used for data analysis and plots.
\end{abstract}

\section{Introduction}

The inflation theory proposes a period of extremely rapid (exponential) expansion of the universe during the very early stage of the universe. although inflationary cosmology has successfully complemented the Standard Model, the process of inflation, in particular its origin, is still largely unknown. Over the past 35 years, numerous models of inflationary expansion of the universe have been proposed. The cosmological inflation is a process in which the size of the universe has increased exponentially at least ${e^{60}} \approx {10^{26}}$ times. The simplest model of inflation is based on the existence of a single scalar field called inflaton. It is possible to build models with other types of fields, for example vector fields, however they have been less used \cite{a. H. Guth,a. D. Linde}. One of the most important ways to test inflationary cosmological models is to compare the computed with the measured values of the observational parameters: the scalar spectral index (${n_s}$) and the tensor-to-scalar power ratio ($r$) \cite{P. a. R. ade et al.,Y. akrami et al.}. Although the analytical methods for approximate calculation of observation parameters is well known there is a drawback: this procedure is often not possible to use for complex models, with a complicated form of the potential, described by a non-standard Lagrangian \cite{D. D. Dimitrijevic,G. S. Djordjevic}.

In this paper we present a quite general method for numerical calculation of the Hubble hierarchy parameters ${\varepsilon _i}$ and the observational parameters of inflation that can be used to validate inflationary models. Using this method, the observational parameters for an inflationary model with a tachyon scalar field, described by a non-standard Lagrangian of the Dirac-Born-Infeld (DBI) type, were calculated. The role of the tachyon field in a braneworld cosmology (Randall-Sundrum II model) was subsequently analyzed \cite{N. Bilic et al.,N. Bilic S. Domazet and G. S. Djordjevic}.

\section{Inflationary models}

The dynamics of a classical real scalar field $(\varphi )$ minimally coupled with gravity is given by
\begin{equation}
S =  - \frac{1}{{16\pi G}}\int {\sqrt { - g} } R{d^4}x + \int {\sqrt { - g} } {\cal L}(X,\varphi ){d^4}x,
\end{equation}	
where $G$ is gravitational constant, $R$ is the Ricci scalar, $g$ is the determinant of the matric tensor, and ${\cal L}(X,\varphi )$ is the Lagrangian, with kinetic term $X \equiv \frac{1}{2}{\partial _\mu }\varphi {\partial _\nu }\varphi$. We will assume the spatially 4-dimensional flat space-time with the standard metric
\begin{equation}
d{s^2} = {g_{\mu \nu }}d{x^\mu }d{x^\nu } = d{t^2} - {a^2}(t)(d{r^2} + {r^2}d{\Omega ^2}).
\end{equation}

In a general case the Lagrangian of a scalar field can be written as an arbitrary function of a scalar $\varphi $ field and a kinetic term $X$. Based on the form of the Lagrangian, several types can be distinguished \cite{S. Li and a. R. Liddle}. 

\subsection{Tachyon inflation}

The software we developed can be used for a wide range of different inflationary models. In this paper we will mainly discuss the models in which inflation is driven by a tachyon field $(\theta )$ originating in the string theory, described by the Lagrangian of the  DBI type and the corresponding Hamiltonian \cite{a. Sen,a. Sen ann. Henri}
\begin{eqnarray}
p &\equiv& {\cal L}(X,\theta ) =  - V(\theta )\sqrt {1 - 2X}  =  - V(\theta )\sqrt {1 - {{\dot \theta }^2}} ,\nonumber\\
\rho  &\equiv& {\cal H} = \frac{{V(\theta )}}{{\sqrt {1 - {{\dot \theta }^2}} }},\label{p and rho}
\end{eqnarray}
where $V(\theta )$ is a tachyon potential which satisfies the following properties \cite{D. Steer and F. Vernizzi,M. Fairbairn and M. H. . Tytgat}
\begin{equation}
V(0) = {\rm{const}},\;\;\;{\kern 1pt} V'(\theta  > 0) < 0,\;\;\;{\kern 1pt} V(|\theta | \to \infty ) \to 0.
\end{equation}

In this case Friedman equation takes the form
\begin{equation}
{H^2} \equiv {\left( {\frac{{\dot a}}{a}} \right)^2} = \frac{{8\pi G}}{3}{\cal H} = \frac{{8\pi G}}{3}\frac{{V(\theta )}}{{\sqrt {1 - {{\dot \theta }^2}} }},\label{Friedman equation}
\end{equation}
where $H$ is the Hubble expansion rate and $a$ is the scale factor.

Dynamics of inflation can be described in the two equivalent ways: using the energy-momentum conservation equation or Hamilton's equations. For a tachyonic potential $V(\theta)$ the energy-momentum conservation equation reads
\begin{equation}
\dot \rho  =  - 3H(P + \rho )\quad \quad  \Rightarrow \quad \quad \frac{{\ddot \theta }}{{1 - {{\dot \theta }^2}}} + 3H\dot \theta  + \frac{1}{V}\frac{{\partial V}}{{\partial \theta }} = 0,\label{equation of motion}
\end{equation}
where is $V' = \partial V/\partial \theta .$

Instead of the energy-momentum conservation equation, the system can be also described by the Hamilton’s equations
\begin{eqnarray}
\dot\theta &=& \frac{{\partial {\cal H}}}{{\partial {\pi _\theta }}},\nonumber\\
\dot\pi_\theta+3H\pi _\theta&=&-\frac{{\partial {\cal H}}}{{\partial \theta }}\label{Hamiltons equations 1}
\end{eqnarray}
where ${\pi _\theta }$ is the conjugate momentum and the Hamiltonian
${\cal H} = \dot \theta {\pi _\theta } - {\cal L}$
is given by Eq. (\ref{p and rho}).

\subsection{The second Randall-Sundrum model}

The Randall-Sundrum (RS) model was originally proposed to solve the hierarchy problem \cite{L. Randall and R. Sundrum 1}, and later it was realized that this model, as well as any similar braneworld model, may have interesting cosmological implications. The second RS model (RSII model) \cite{L. Randall and R. Sundrum 2} describes a $4+1$ dimensional anti de Sitter ($\text{AdS}_5$) universe containing two 3-branes with opposite tensions, separated in the fifth dimension, with observers on the positive tension brane. The fluctuation of the interbrane distance along the extra dimension implies the existence of the radion $\varphi$, a massless scalar field that causes a distortion of the bulk geometry.

The total action, as seen on the observer’s brane, is \cite{N. Bilic et al.,N. Bilic and G. B. Tupper}
\begin{eqnarray}
S &=& \int {{d^4}} x\sqrt { - g} \left( { - \frac{R}{{16\pi G}} + \frac{1}{2}{g^{\mu \nu }}{\varphi _{,\mu }}{\varphi _{,\nu }}} \right)\nonumber\\ 
&-& \int {{d^4}} x\sqrt { - g} {\mkern 1mu} \frac{\sigma }{{{k^4}{\theta ^4}}}{(1 + {k^2}{\theta ^2}\eta )^2}\sqrt {1 - \frac{{{g^{\mu \nu }}{\theta _{,\mu }}{\theta _{,\nu }}}}{{{{(1 + {k^2}{\theta ^2}\eta )}^3}}}} ,
\end{eqnarray}
where the second term is the action of the brane, and $k = 1/l$ is the inverse of $\text{AdS}_5$ curvature radius, $\sigma $ is the brane tension and $\eta$ is the rescaled radion field $\eta  = {\sinh ^2}\left( {\sqrt {4/3\,\pi G} \phi } \right)$. The total Lagrangian and the Hamiltonian for the brane and the radion are
\begin{eqnarray}
{\cal L} &=& \frac{1}{2}{g^{\mu \nu }}{\phi _{,\mu }}{\phi _{,\nu }} - \frac{{\lambda {\psi ^2}}}{{{\theta ^4}}}\sqrt {1 - \frac{{{g^{\mu \nu }}{\theta _{,\mu }}{\theta _{,\nu }}}}{{{\psi ^3}}}},\nonumber\\
{\cal H} &=& \frac{1}{2}{g^{\mu \nu }}{\phi _{,\mu }}{\phi _{,\nu }} + \frac{{\lambda {\psi ^2}}}{{{\theta ^4}}}{\left( {1 - \frac{{{g^{\mu \nu }}{\theta _{,\mu }}{\theta _{,\nu }}}}{{{\psi ^3}}}} \right)^{ - 1/2}},
\end{eqnarray}
where $\psi  = 1 + {k^2}{\theta ^2}\eta$. Following the standard procedure, the Hamilton's equations are
\begin{eqnarray}
\dot \phi  &=& \frac{{\partial {\cal H}}}{{\partial {\pi _\phi }}},\nonumber\\
\dot \theta  &=& \frac{{\partial {\cal H}}}{{\partial {\pi _\theta }}},\nonumber\\
{\dot \pi _\phi } + 3H{\pi _\phi } &=&  - \frac{{\partial {\cal H}}}{{\partial \phi }},\nonumber\\
{\dot \pi _\theta } + 3H{\pi _\theta } &=&  - \frac{{\partial {\cal H}}}{{\partial \theta }},\label{Hamiltons equations 2}
\end{eqnarray}
where $\pi _\phi ^{} = \partial L/\partial \dot \phi$ and $\pi _\theta ^{} = \partial L/\partial \dot \theta$ are the conjugate momenta. In the spatially flat Randall-Sundrum cosmology the Hubble expansion rate $H$ is related to the Hamiltonian via the modified Friedmann equation \cite{R. Maartens and K. Koyama}
\begin{equation}
H \equiv \frac{{\dot a}}{a} = \sqrt {\frac{{8\pi G}}{3}{\cal H}\left( {1 + \frac{{2\pi G}}{{3{k^2}}}{\cal H}} \right)} .\label{modified Friedmann equation}
\end{equation}	

\subsection{The Observational parameters}
Over the past twenty years many research missions have performed numerous observations and collected large amounts of data. The latest and the most important observational results for testing the cosmological models of inflation are provided by the Planck collaboration \cite{P. a. R. ade et al.,Y. akrami et al.}

The inflation slow roll  parameters $({\varepsilon _i})$ \cite{D. Steer and F. Vernizzi,D. J. Schwarz C. a. Terrero-Escalante} are defined as follows
\begin{equation}
\epsilon_0\equiv\frac{H_{*}}{H},\quad \epsilon_i\equiv\frac{d\ln|\epsilon_{i-1}|}{Hdt},\quad  i\geq 1,\label{slow roll  parameters}
\end{equation}
where $H_{*}$ is the Hubble rate at an arbitrarily chosen time, and number of e-folds $(N)$ is 
\begin{equation}
dN(t) = H(t)dt\quad \quad  \Rightarrow \quad \quad N(t) = \int\limits_{{t_{\rm{i}}}}^{{t_{\rm{e}}}} {H(t)dt} ,\label{e-folds}
\end{equation}
where $t_i$ is the beginning and $t_e$ is the end of inflation, ${\varepsilon _i}({t_{\rm{e}}}) = 1.$ 

Now, the observational parameters, the scalar spectral index (${n_s}$) and the tensor-to-scalar power ratio ($r$), in lowest order in the  slow roll parameters are
\begin{equation}
r = 16{\varepsilon _1},\quad \quad {n_s} = 1 - 2{\varepsilon _1} - {\varepsilon _2},
\end{equation}
The observational parameters in the second order in the slow roll parameters are
\begin{equation}
\begin{split}
r&=16\epsilon_{1}(t_{i})\big[1-2\alpha\epsilon_{1}(t_{i})+C\epsilon_{2}(t_{i})\big], 
\\
n_{s}=1-2\epsilon_{1}(t_{i})-2\epsilon_{2}(t_{i})&-\big[2\epsilon^{2}_{i}(t_{i})+(2C+3-2\alpha)\epsilon_{1}(t_{i})\epsilon_{2}(t_{i})+C\epsilon_{2}(t_{i})\epsilon_{3}(t_{i})\big],
\end{split}\label{observational parameters}
\end{equation}
where $C =  - 0.72$, $\alpha  = 1/6$ for tachyon models in the standard cosmology or $\alpha  = 1/12$ in the case of the Randall-Sundrum cosmology.

According to the most restrictive results ({\em Planck TT,TE,EE + lowW + lensing + BK14}) the expected values of observational parameters for the base-$\Lambda$CDM cosmology are \cite{Y. akrami et al.}:
\begin{eqnarray}
n_s &=& 0.9665 \pm 0.0038\quad(68\% CL),\nonumber\\
r_{0.002} &<& 0.064\quad(95\% CL),
\end{eqnarray}

\begin{figure}[t]
\centering
   \begin{subfigure}{0.49\linewidth} \centering
     \includegraphics[width=\linewidth]{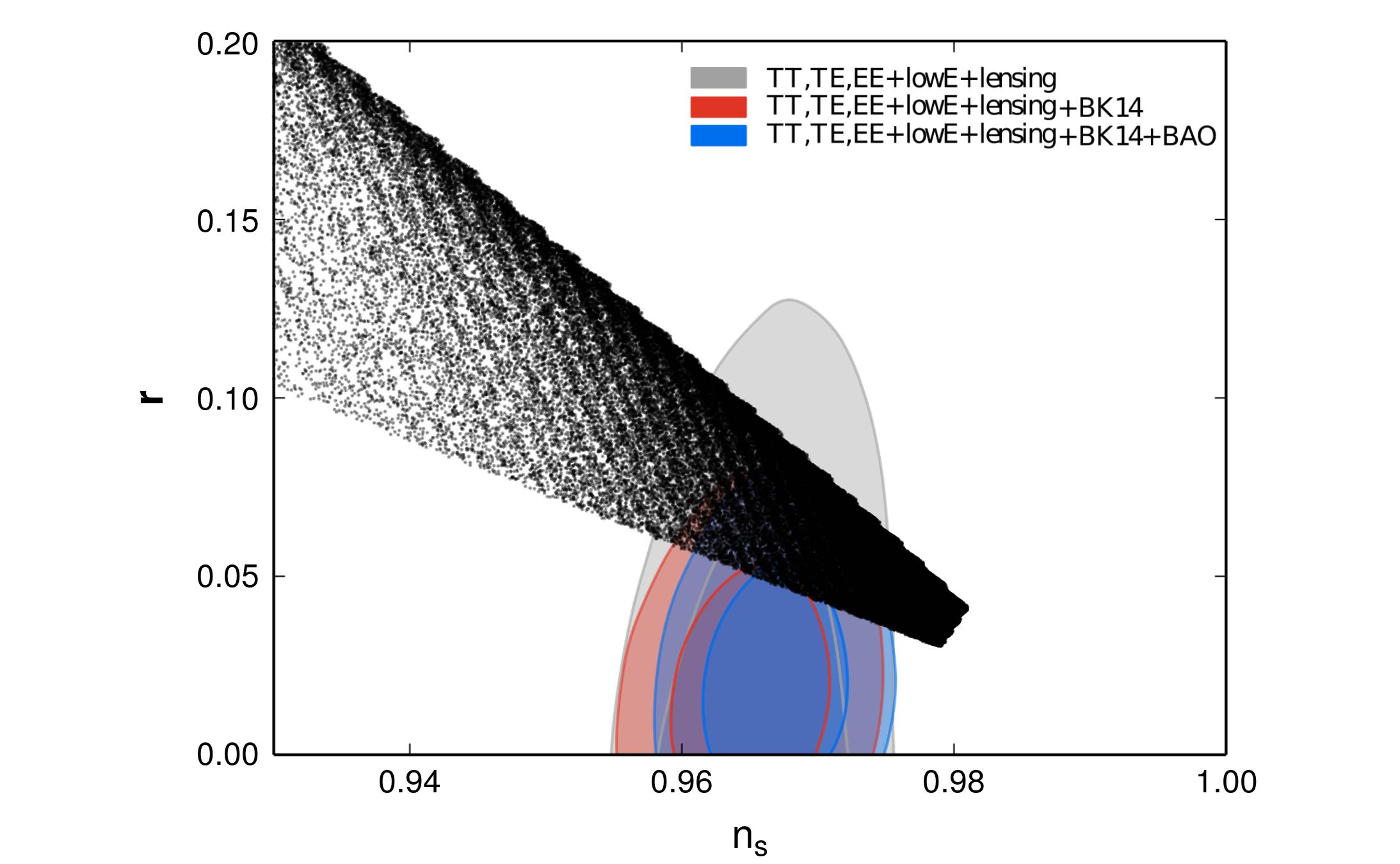}
   \end{subfigure}
   \begin{subfigure}{0.49\linewidth} \centering
     \includegraphics[width=\linewidth]{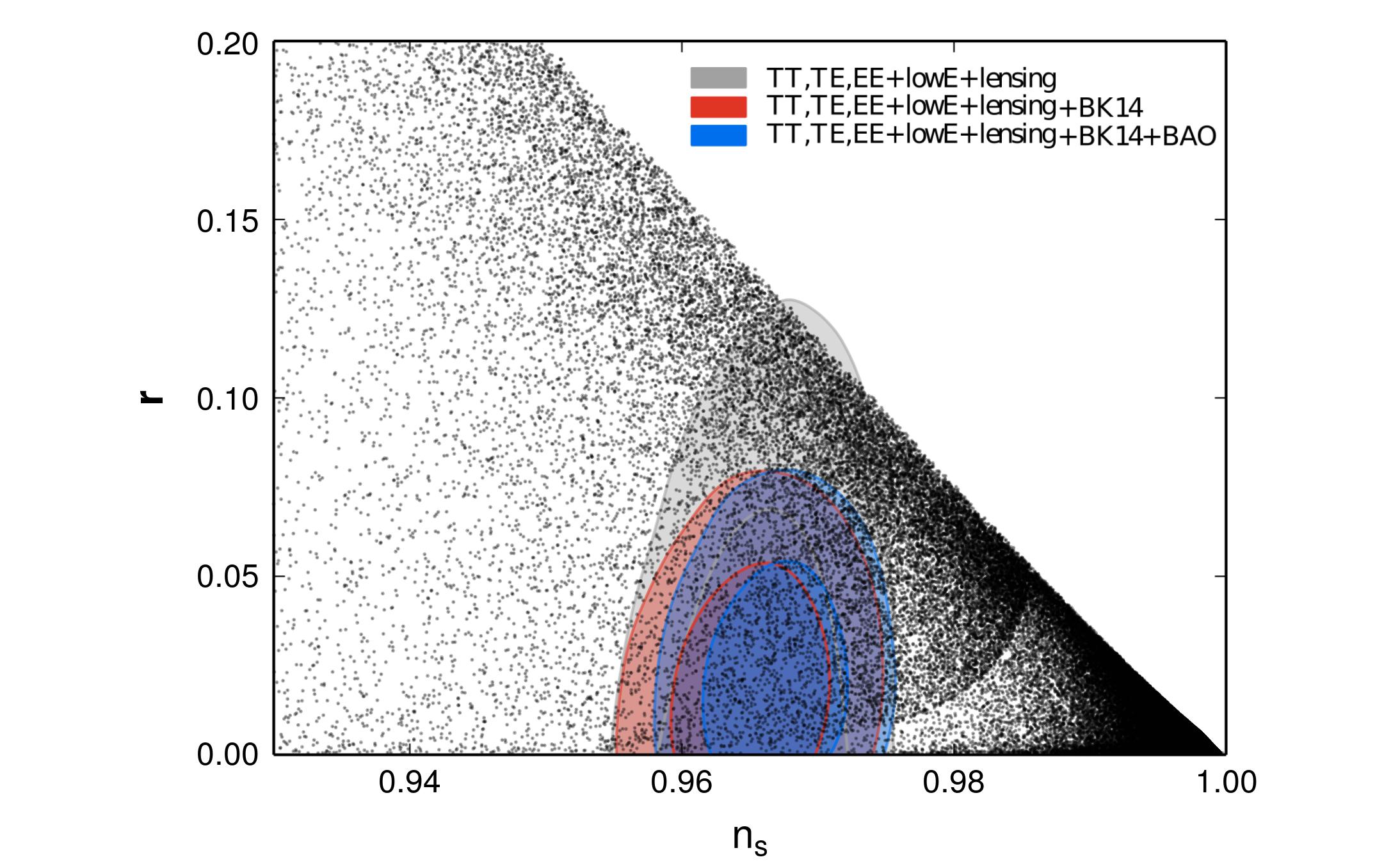}
   \end{subfigure}
\caption{Observational parameters for $V(\theta ) = 1/\cosh (\theta )$ in the standard cosmology (SC) (\ref{Friedman equation}) (left) and the Randall-Sundrum cosmology (\ref{modified Friedmann equation}) (right) for various $N$ and $\kappa$ chosen randomly, $30 \le N \le 150$, $0 \le \kappa \le 15$. The results are compared with the observational constraints from Planck collaboration \cite{Y. akrami et al.}.} \label{fig:one}
\end{figure}

\begin{figure}[t]
\centering
   \begin{subfigure}{0.49\linewidth} \centering
     \includegraphics[width=\linewidth]{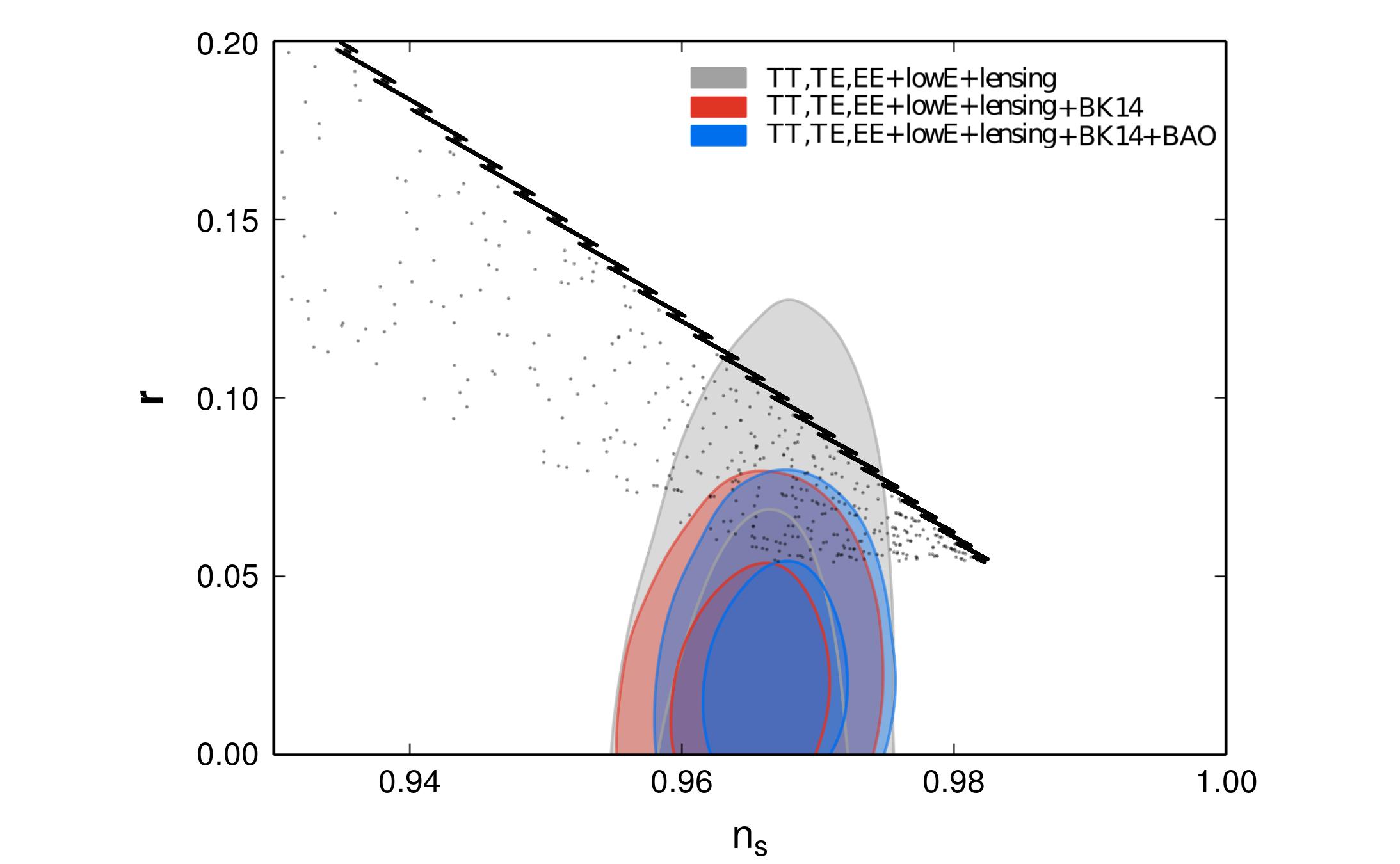}
   \end{subfigure}
   \begin{subfigure}{0.49\linewidth} \centering
     \includegraphics[width=\linewidth]{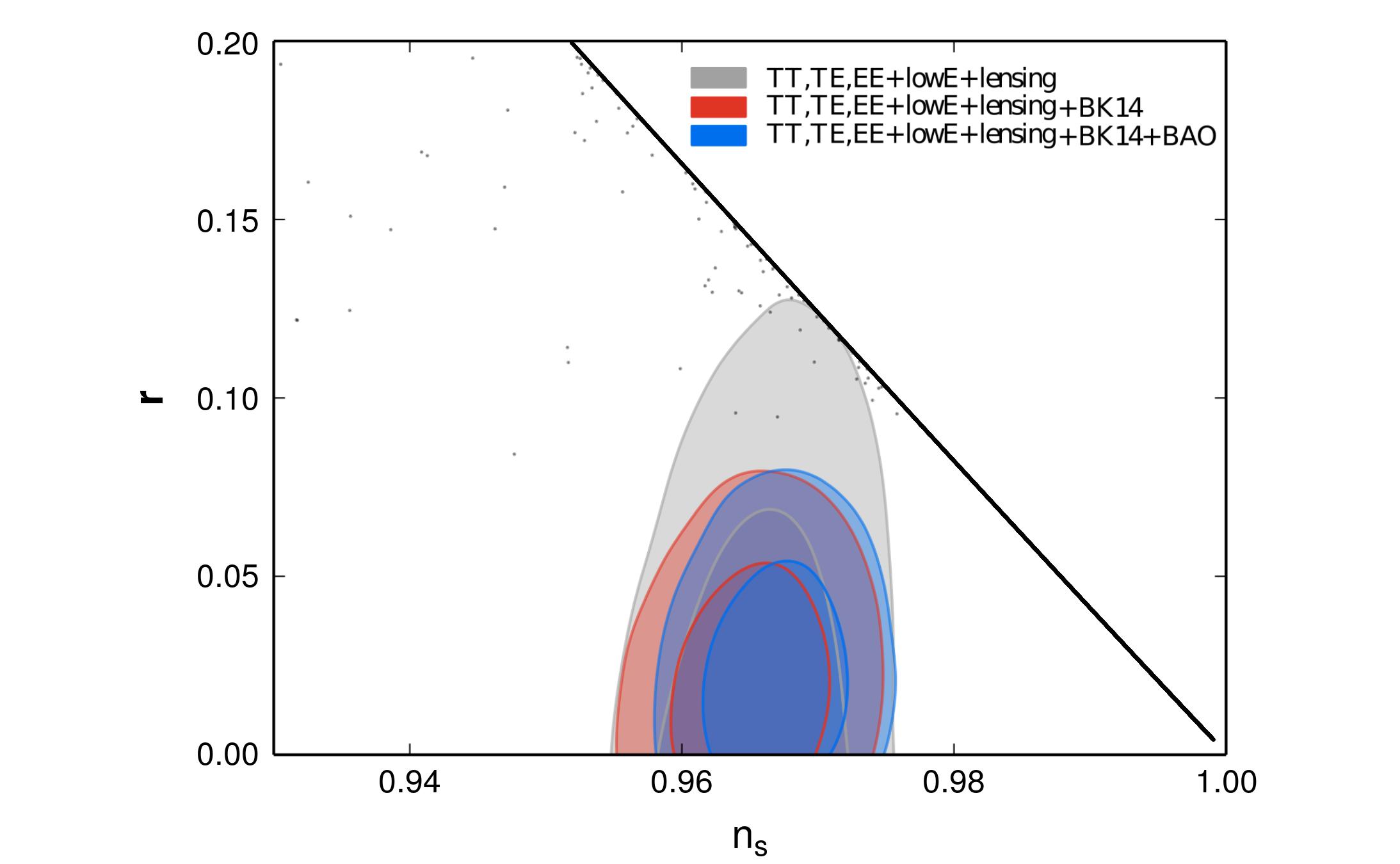}
   \end{subfigure}
\caption{Observational parameters for $V(\theta ) = \exp ( - \theta )$ for various $N$ and $\kappa$ chosen randomly, $30 \le N \le 150$, $0 \le \kappa \le 15$. Friedman equation is given for SC by (\ref{Friedman equation}) (left) and for RS by (\ref{modified Friedmann equation}) (right).} \label{fig:two}
\end{figure}

\begin{figure}[t]
\centering
   \begin{subfigure}{0.49\linewidth} \centering
     \includegraphics[width=\linewidth]{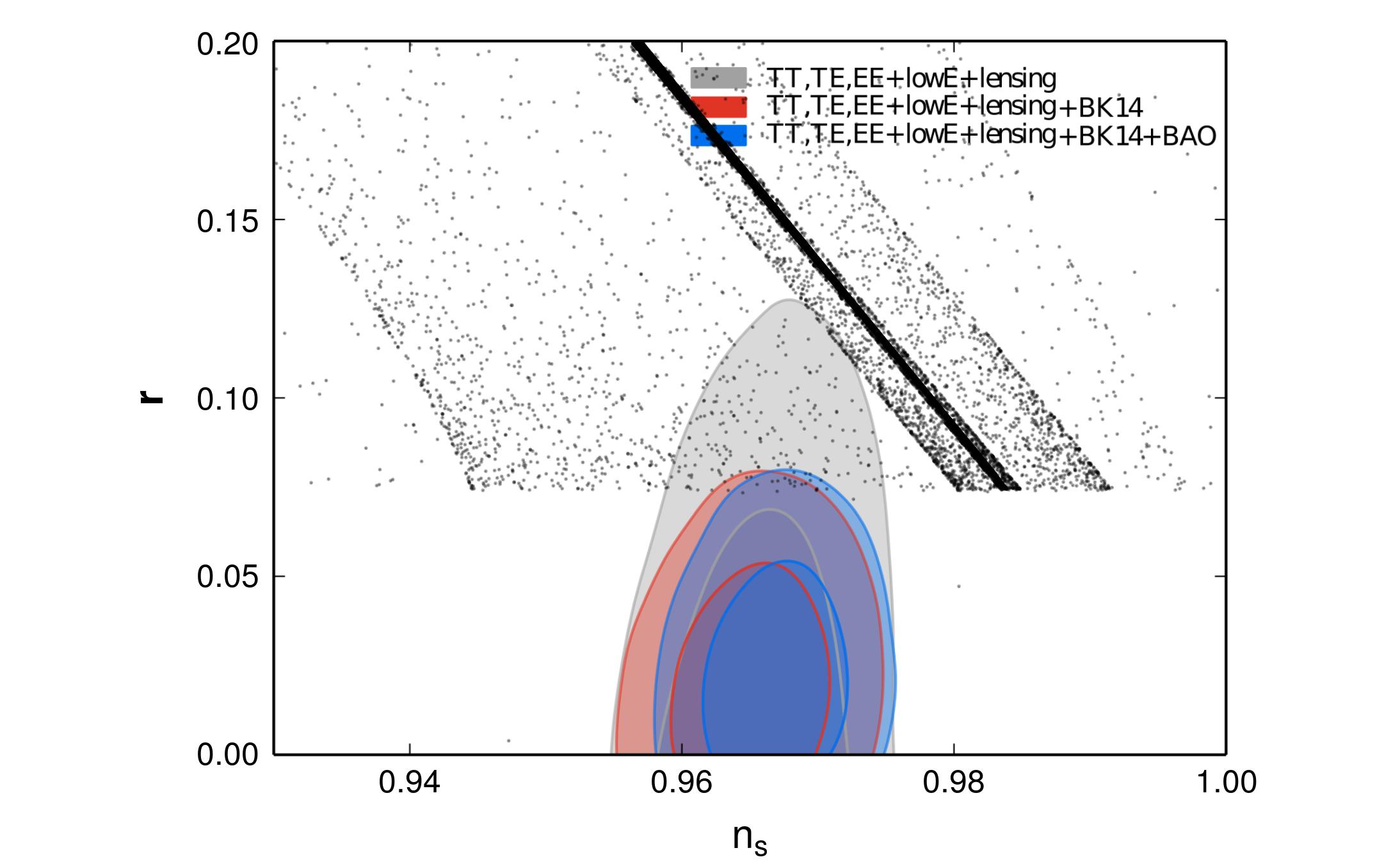}
   \end{subfigure}
   \begin{subfigure}{0.49\linewidth} \centering
     \includegraphics[width=\linewidth]{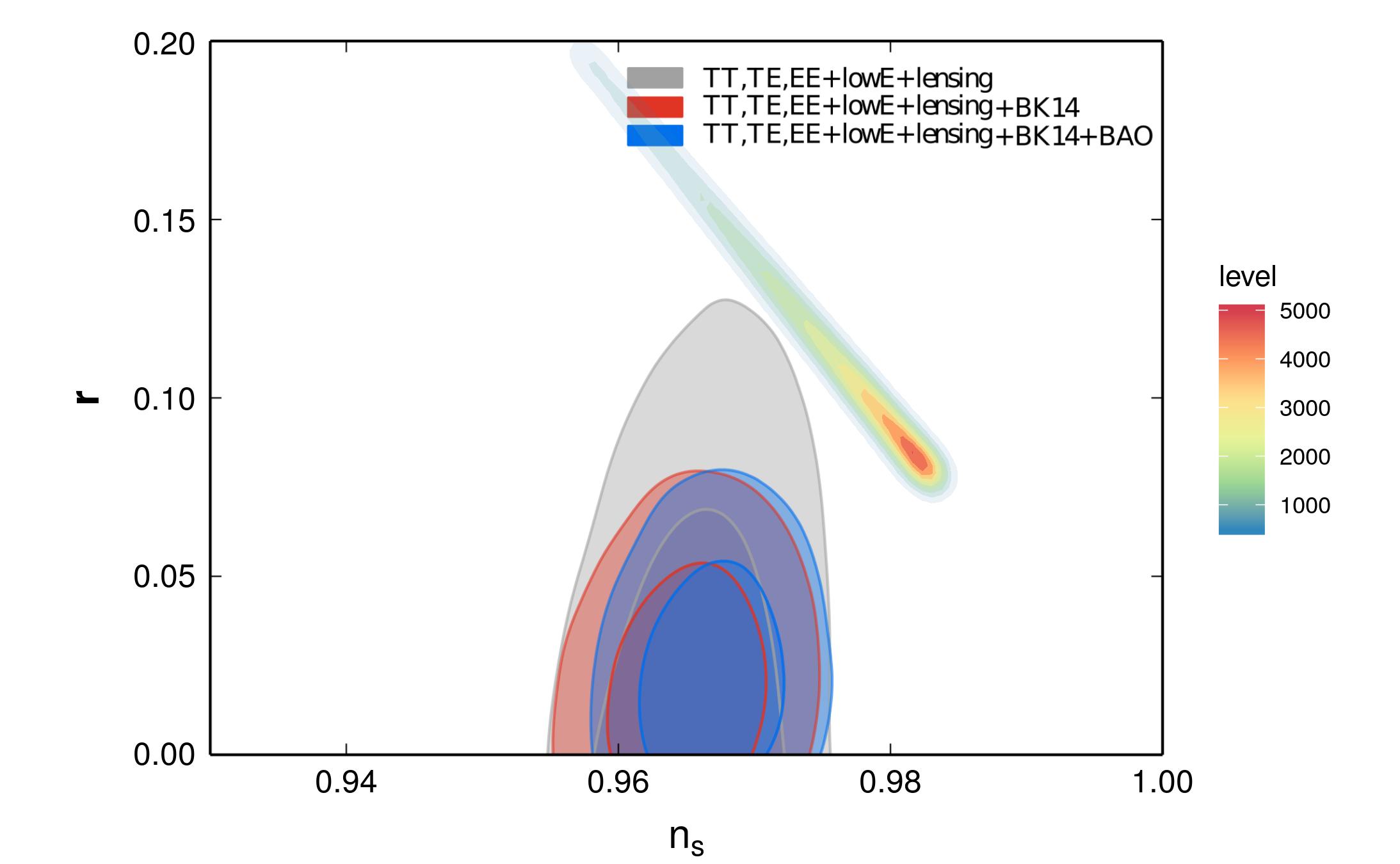}
   \end{subfigure}
\caption{Observational parameters for $V(\theta ) = {\theta ^{ - 4}}$, the Friedman equation is given by (\ref{modified Friedmann equation}) for various $N$ and $\kappa$ chosen randomly,
$30 \le N \le 150$, $0 \le \kappa \le 15$. Each dot in the plot on the left side represents the results for chosen $(N,\kappa)$. The same results are shown in the plot on the right side, the color represents density of the results: red – higher, blue - lower density.} \label{fig:three}
\end{figure}

\begin{figure}[t]
\centering
\includegraphics[width=0.5\linewidth]{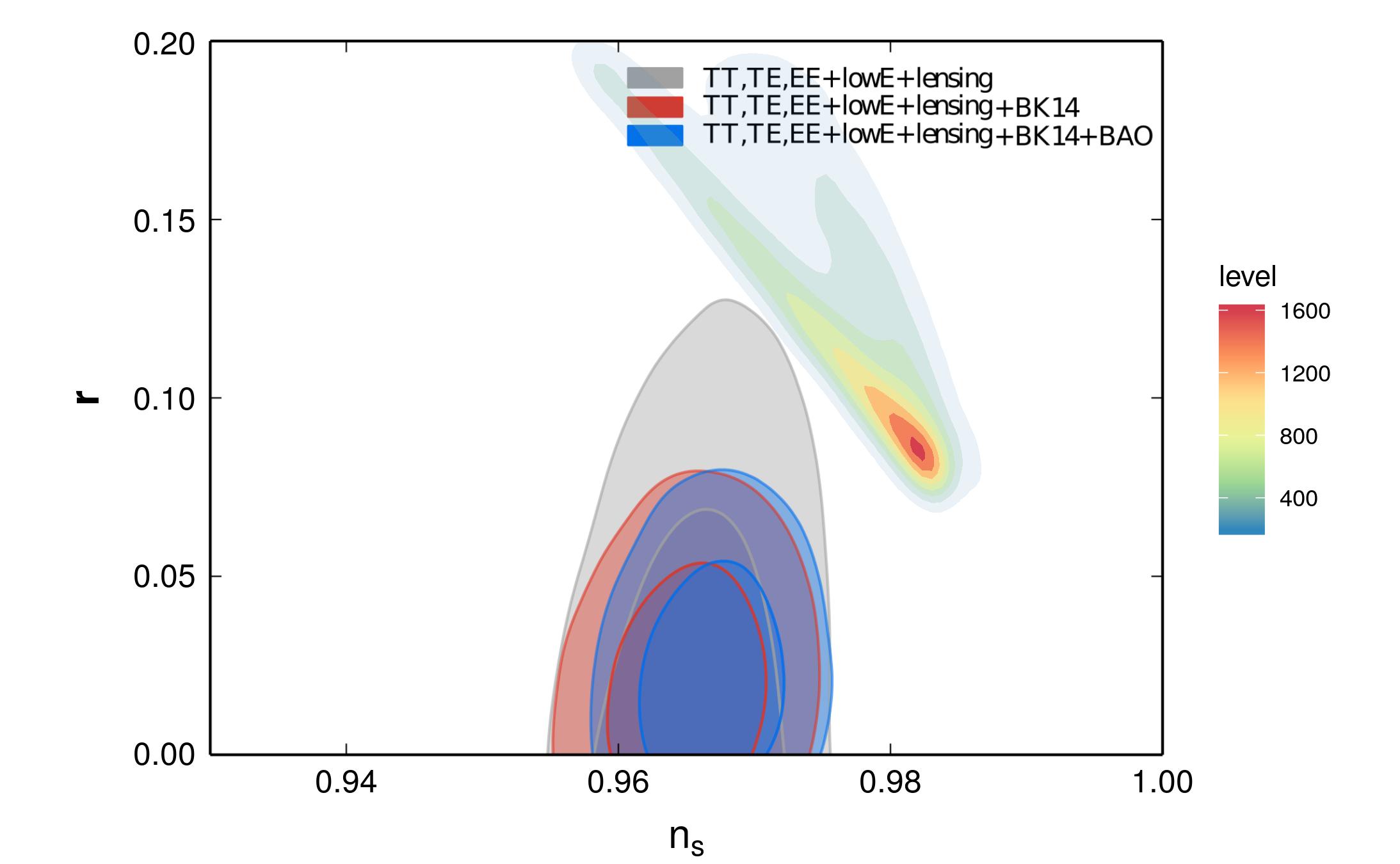}
\caption{Results for observational parameters in $(n_s, r)$ plane. The color represents a density of the dots obtained for the RSII model (right). About 100,000 observational parameters were calculated for randomly chosen $N$, $\kappa$ and ${\phi _0}$ in the range
$30 \le N \le 150$, $0 \le \kappa \le 15$ and $0 < {\phi _0} < 1$. } \label{fig:four}
\end{figure}

The results we have calculated for different tachyon potentials and the RSII model are compared with the Planck results in Fig. \ref{fig:one} - \ref{fig:four}. 

\section{Numerical results} 
The already mentioned software we developed for numerical calculations of the observational parameters (\ref{slow roll  parameters})-(\ref{observational parameters}) is based on the standard analytical procedure in the slow-roll regime $(\dot \theta  \ll 1,{\rm{ }}\ddot \theta \ll 3H\dot \theta ).$ Here the slow-roll approximation is used only to determinate the initial conditions, and evolution of the system is calculated numerically.

After nondimensionalization \cite{N. Bilic et al.} of the equation of motion (\ref{equation of motion}) or of the system of the Hamilton’s equations (\ref{Hamiltons equations 1}) and (\ref{Hamiltons equations 2}) for tachyon and the RSII model, respectively, it is possible to obtain a system of first order differential equations with only one free parameter $(\kappa ).$

The system is evolved numerically using Runge-Kutta method starting from $t = 0$ up to some large $t_max$ of the order of $100l$. The function $N(t)$ is solved simultaneously using (\ref{e-folds}) with the initial condition $N(0) = 0$. The time evolution of the Hubble hierarchy parameters are obtained using (\ref{slow roll  parameters}). The end of inflation $({t_{\rm{e}}})$ is calculated as the solution of the equation ${\varepsilon _i}({t_{\rm{e}}}) = 1.$ If the inflation does not end in the chosen time interval the order of $t_{max}$ is raised and calculation is repeated. The beginning of inflation $t_i$ is then obtained by requiring $N({t_{\rm{e}}}){\rm{ }} - {\rm{ }}N({t_{\rm{i}}}){\rm{ }} = {\rm{ }}N$. The simulation was repeated (about 100,000 times) for each model for randomly selected combination of $N$ and $\kappa$ in a chosen interval ($30 \le N \le 150$, $0 \le \kappa \le 15$), and results are analyzed and plotted in R language \cite{R Development Core Team}.

Evidently, a comparison of the computed results with Planck data shows a reasonable agreement. A comparison of the presented results with the results published in \cite{N. Bilic et al.,M. Milosevic et al.,N. Bilic et al. in aIP}, obtained in a different way, or using the previous version of the program, gives also a good overlap.

\begin{figure}[t]
\centering
   \begin{subfigure}{0.49\linewidth} \centering
     \includegraphics[width=\linewidth]{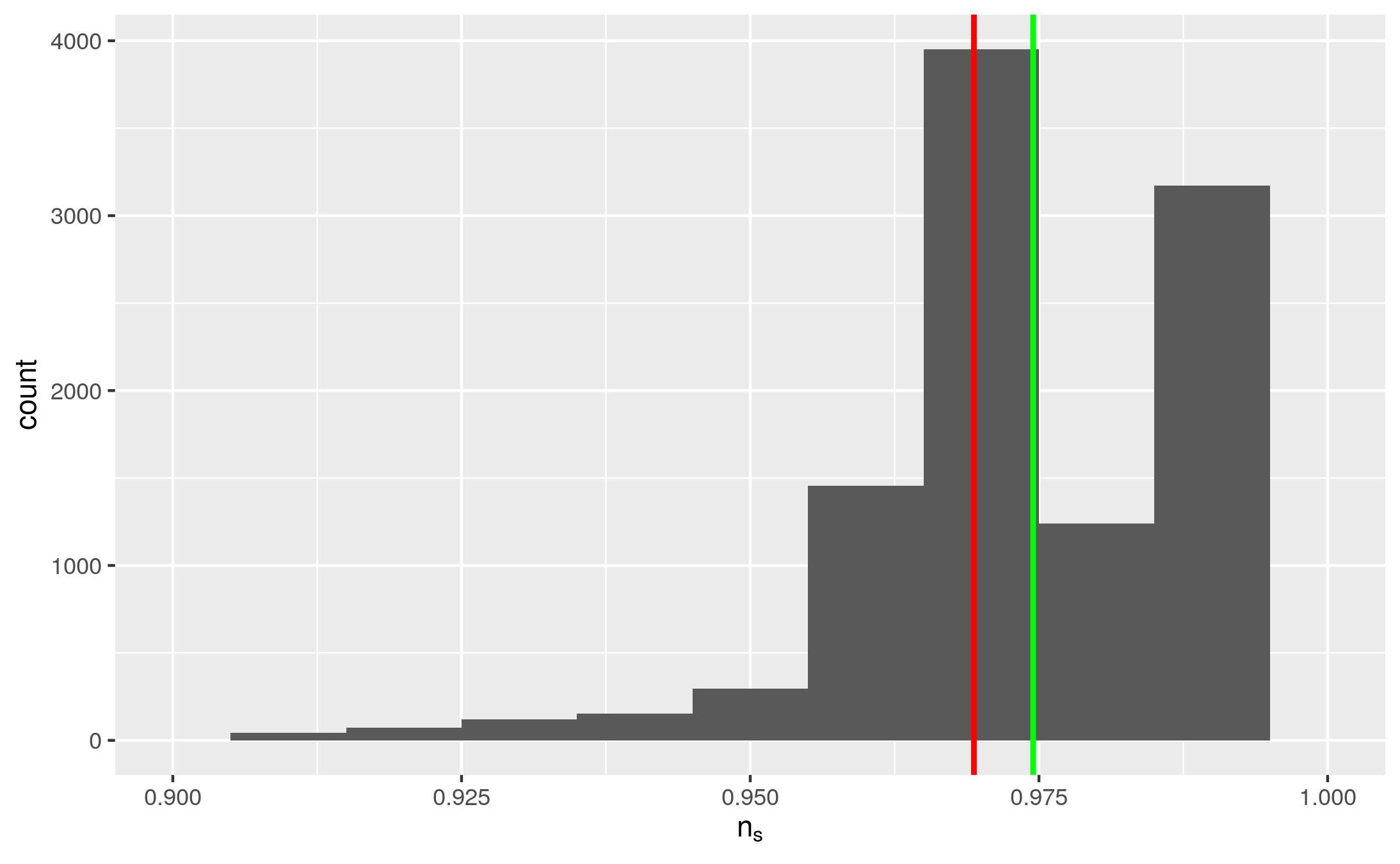}
   \end{subfigure}
   \begin{subfigure}{0.49\linewidth} \centering
     \includegraphics[width=\linewidth]{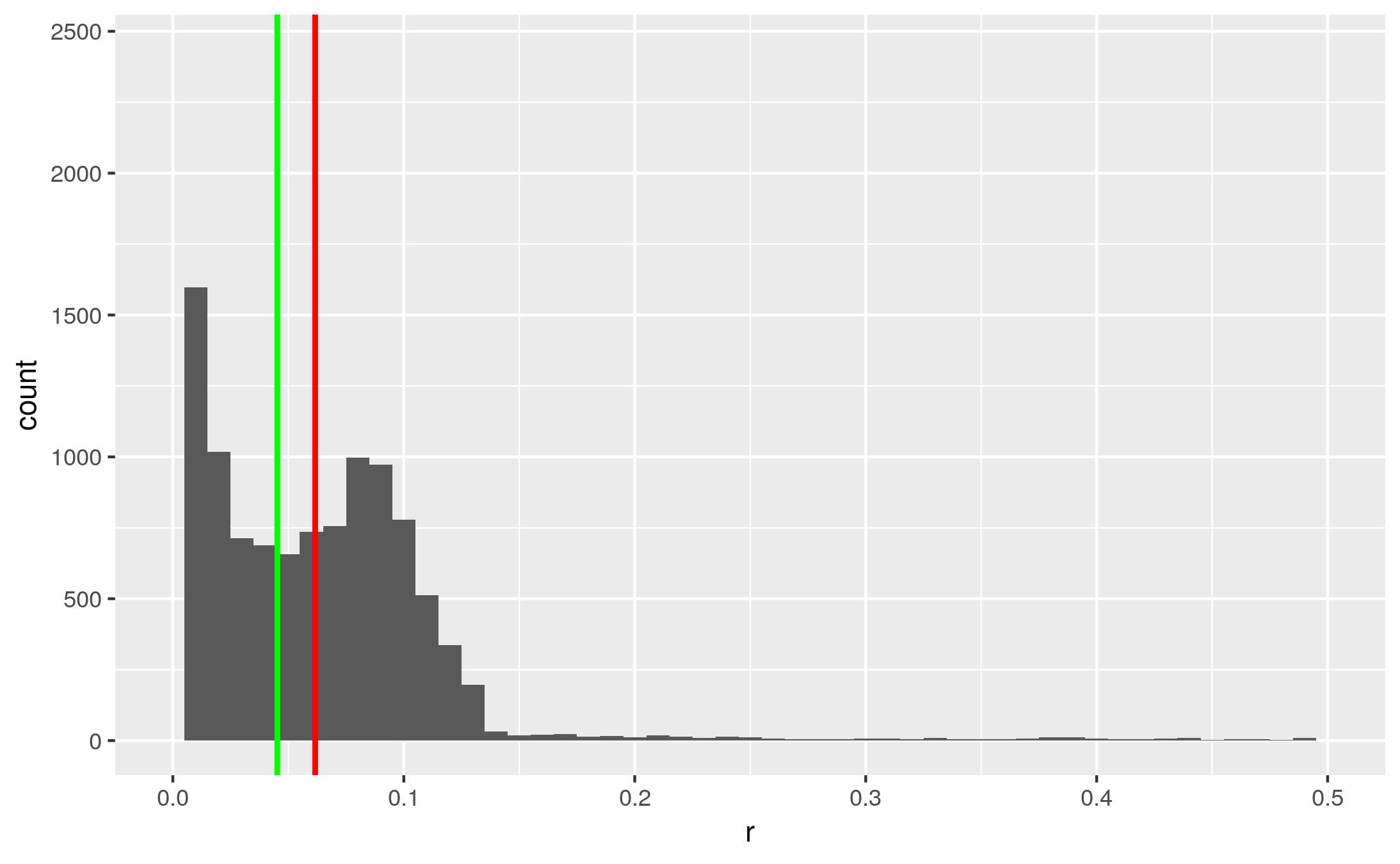}
   \end{subfigure}
\caption{Statistical distribution of the scalar spectral index $n_s$ (left) and the tensor-to-scalar ratio $r$ (right) for the $V(\theta ) = 1/\cosh (\theta )$ in Randall-Sundrum cosmology (\ref{modified Friedmann equation}). The free parameters are in the range $30 \le N \le 150$, $0 \le \kappa \le 15$. The solid vertical lines in the histograms mark the mean $\mu ({n_s}) = {\rm{0}}{\rm{.9694}}$ and $\mu (r) = 0.062$ (red) and median ${\tilde n_s} = 0.9745$ and $\tilde r = 0.045$ (green).} \label{fig:five}
\end{figure}

The best agreement with observational data provided by Planck collaboration was achieved for the potential $V(\theta ) = 1/\cosh (\theta )$ and the free parameters in the interval  $60 \le N \le 90$, $1 \le \kappa \le 10$ for the Randall-Sundrum cosmology (11). The statistical distributions of computed results are shown in the histograms (Fig. \ref{fig:five}), mean $\mu ({n_s}) = {\rm{0}}{\rm{.9694}}$, $\mu (r) = 0.062$, and median ${\tilde n_s} = 0.9745$, $\tilde r = 0.045$.

We would like to stress that our program provides fast numerical calculation of the observational parameters for known models.
\section{Conclusion}
The program we developed and used here has been applied to a limited set of models, mainly to the pure tachyonic and the RSII inflationary cosmological models. The theoretical input is provided by the Hamilton and the Friedmann equations for the chosen potentials and corresponding parameters. The program can readily be used for a much wider set of the models. To apply the program to a new model one must determine its corresponding equations and include these equations in the program only.

The next steps are: to extend the program to be applicable for new and different types of inflationary models, to improve the program in such a way that only the Hamiltonian or the Lagrangian of a model are their inputs. The corresponding system of differential equations would be determined by symbolic computation. after these improvements the program will be published as a free software followed by appropriate documentation.

Finally, the best fitting result is obtained for $V(\theta ) = 1/\cosh (\theta )$. It opens good opportunity for further research based on this potential in the context of the RSII model and the holographic cosmology. These results will be published elsewhere. 

\section*{Acknowledgments}

This work has been supported by ICTP - SEENET-MTP project NT-03: Cosmology - Classical and Quantum Challenges. The work of N. Bili\' c. has been supported by the H2020 CSa Twinning project No. 692194, “RBI-T-WINNING.” and partially supported by the STSM CaNTaTa-COST grant. D. D. Dimitrijevi\' c, G. S. Djordjevi\' c, M. Milo\v sevi\' c and M. Stojanovi\' c acknowledge support provided by the Serbian Ministry for Education, Science and Technological Development under the projects No. 176021 (G.S.Dj, M.M), No. 174020 (D.D.D) and 176003 (M.S). G. S. Djordjevic is grateful to the CERN-TH Department for hospitality and support.

\end{document}